# Direct band gap and strong Rashba effect in van der Waals heterostructures of InSe and Sb single layers

Dangqi Fang[1,2,*], Siyu Chen[2], Yaqi Li[1], and Bartomeu Monserrat[2,3,†]

[1]MOE Key Laboratory for Nonequilibrium Synthesis and Modulation of Condensed Matter, School of Physics, Xi'an Jiaotong University, Xi'an 710049, China

[2]Cavendish Laboratory, University of Cambridge, J. J. Thomson Avenue, Cambridge CB3 0HE, United Kingdom

[3]Department of Materials Science and Metallurgy, University of Cambridge, 27 Charles Babbage Road, Cambridge CB3 0FS, United Kingdom

Abstract

Van der Waals heterostructures formed by stacking different types of 2D materials are attracting increasing attention due to new emergent physical properties such as interlayer excitons. Recently synthesized atomically thin indium selenide (InSe) and antimony (Sb) individually exhibit interesting electronic properties such as high electron mobility in the former and high hole mobility in the latter. In this work, we present a first-principles investigation on the stability and electronic properties of ultrathin bilayer heterostructures composed of InSe and Sb single layers. The calculated electronic band structures reveal a direct band gap semiconducting nature of the InSe/Sb heterostructures independent of stacking pattern. Taking spin-orbit coupling into account, we find a large Rashba spin splitting at the bottom of conduction band, which originates from the atomic spin-orbit coupling with the symmetry breaking in the heterostructure. The strength of the Rashba spin splitting can be tuned by applying in-plane biaxial strain or an out-of-plane external electric field. The presence of large Rashba spin splitting together with a suitable band gap in InSe/Sb bilayer heterostructures make them promising candidates for spin field-effect transistor and optoelectronic device applications.

*fangdqphy@xjtu.edu.cn

†bm418@cam.ac.uk

# I. Introduction

Indium selenide (InSe) is a group III-VI layered metal chalcogenide semiconductor[1]. Each of its layers possesses a honeycomb lattice that comprises four covalently bonded Se-In-In-Se atomic planes and the layers are interconnected through van der Waals (vdW) forces. Two dimensional (2D) atomically thin InSe has recently received a lot of attention due to its exotic electronic and optical properties. The valence band of InSe exhibits an inverted "Mexican hat" shape in the few layers limit but evolves into a conventional parabolic shape with increasing thickness, as initially predicted theoretically [2,3] and recently confirmed by experiments[4–6]. There are several reports on 2D InSe made by mechanical exfoliation[6–9], molecular beam epitaxy[5], or chemical vapor transport methods[10]. The electron mobility in 2D InSe has been reported to exceed $10^3$ $cm^2V^{-1}S^{-1}$ at room temperature, making this material suitable for high-performance electronic device applications[7]. Moreover, the quantum Hall effect has been observed in high quality chemically synthesized InSe[10]. 2D InSe shows great promise not only for charge-based devices, but also for spin-based devices due to the spin-orbit coupling (SOC). Premasiri *et al.* observed an electric-field tunable Rashba SOC effect in gated multilayer InSe[9]. Zhou *et al.* predicted that perpendicular electric fields lead to a tunable current-induced spin polarization in monolayer InSe[11].

Single-layer antimony (Sb), named antimonene, with a buckled honeycomb structure has been shown theoretically to have an indirect band gap[12–14] and a high hole mobility of about 1330 $cm^2V^{-1}S^{-1}$ at room temperature[15]. The synthesis of antimonene has recently been realized in experiments based on both epitaxial growth[16] and liquid-phase exfoliation[17].

The physical properties of 2D materials become even richer when considering that they can be vertically stacked to form bilayer or superlattice vdW heterostructures[18]. An interface transition-induced broad optical spectrum is reported in vdW heterostructures such as InSe/WS$_2$[19] and GaTe/InSe[20]. Proximity between 2D magnets and other layered materials offers controllable ways to tailor the electronic structure of adjacent layers[21,22]. For example, in recent low temperature experiments on CrI$_3$/WSe$_2$ vdW heterostructures, the photoluminescence of monolayer WSe$_2$ shows a valley splitting of about 3.5 meV, equivalent to an effective magnetic field of nearly 13 T, due to the strong magnetic exchange field between CrI$_3$ spins and WSe$_2$ excitons[21]. Rashba spin splitting has also been predicted in vdW heterostructures such as MoS$_2$/Bi(111)[23] and GaX/MX$_2$ (M=Mo, W; X=S, Se, Te)[24].

The Rashba effect plays a vital role in the electronic states of heterostructures and surfaces. The 2D electron gas Hamiltonian with Rashba SOC is of the form[25]:

$$H_R = \frac{\hbar^2 \vec{k}_\parallel^2}{2m^*} + \alpha_R \vec{\sigma} \cdot (\vec{k}_\parallel \times \hat{z}), \quad (1)$$

where $\vec{k}_\parallel = (k_x, k_y, 0)$ is the in-plane momentum of the electrons, $\alpha_R$ is the Rashba parameter that is used to characterize the strength of the Rashba effect, $\vec{\sigma} = (\sigma_x, \sigma_y, \sigma_z)$ is the Pauli matrix vector, and $\hat{z}$ is the out-of-plane unit vector. Rashba bands exhibit spin-momentum locking that can be used to manipulate the spin direction by means of an electric field[26] and is therefore potentially useful for applications such as the spin field-effect transistor[27]. It has been experimentally observed at heavy metal surfaces[28,29], in semiconductor quantum wells[30], and in the bulk polar semiconductor BiTeI[31], and theoretically predicted in several 2D materials such as thin-film LaOBiS$_2$[32], monolayer BiSb[33], and polar transition-metal dichalcogenide monolayers[34]. It is worth noting that a different interpretation for the Rashba effect has been proposed, which attributes the Rashba-type band splitting to the formation of an asymmetric charge distribution from the local orbital angular momentum and crystal momentum[35,36].

Given the small lattice mismatch (less than 1% according to our calculations) between InSe and Sb single layers and the strong SOC in these two materials, it is interesting to explore InSe/Sb vdW heterostructures and to investigate how the interlayer interaction affects their electronic properties. A very recent theoretical study of an InSe/Sb heterostructure finds that the band gap can be tuned by applying vertical strain or an external electric field[37], but a detailed study about the role of SOC is so far missing. In this work, we systematically investigate the stacking geometry and electronic band structure of InSe/Sb bilayer heterostructures. We demonstrate the direct band gap semiconducting nature and large Rashba spin splitting at the conduction band edge around the Γ point, rendering this material potentially suitable for low temperature spintronic and optoelectronic applications.

## II. Computational details

Our first-principles calculations are based on density functional theory using the projector augmented-wave method as implemented in the Vienna *ab initio* simulation package (VASP)[38–40]. The exchange-correlation functional is described within the generalized gradient approximation (GGA) as parameterized by Perdew, Burke, and Ernzerhof (PBE)[41]. The following electrons are treated as valence states: In($4d^{10}5s^25p^1$), Se($4s^24p^4$), and Sb($5s^25p^3$). The cut-off energy for the plane wave basis set is chosen to be 500 eV. A vacuum region of more than 12 Å in the direction perpendicular to the plane of 2D materials is employed to suppress the interaction between neighboring periodic images. The Becke-Johnson damping of the DFT-D3 method is used to take into account the vdW interaction[42]. The Brillouin-zone samplings of Γ-centered 10×10×1 and 20×20×1 **k**-point grids are used for structural relaxations and density of states calculations, respectively[43]. All atoms are relaxed until the Hellmann-Feynman forces are less than 0.005 eV/Å and the in-plane lattice parameters are also relaxed to determine their optimal values. Our calculations show that the lattice parameters for bulk γ-InSe optimized using the DFT-D3 method are a = 4.037 Å and c = 24.914 Å, which are in good agreement with the experimental values of a = 4.002 Å and c = 24.946 Å[1]. Additionally, the DFT-D3 method has been successfully used to investigate the properties of vdW heterostructures such as the InSe/SiGe bilayer[44]. We have checked that SOC has a negligible effect on the structural relaxations, so SOC is only included in the band structure calculations based on the relaxed geometries[45]. To compensate for the band gap underestimation of the PBE functional, we carry out additional Heyd-Scuseria-Ernzerhof (HSE06) screened hybrid functional calculations[46]. The PYPROCAR[47] and VESTA[48] codes are used to plot the spin texture and the atomic models, respectively.

# III. Results and discussion

We first investigate the properties of the separate InSe and Sb single layers. The optimized lattice parameters of InSe and Sb single layers are 4.026 Å and 4.064 Å, respectively, which are in good agreement with previous calculations[44,49]. Their band structures and band gaps calculated without and with SOC are presented in Fig. 1 and Table I. The calculations without SOC show that both materials are indirect-band-gap semiconductors. The single layer of InSe exhibits a peculiar "Mexican hat"-shaped valence band. Its band gap is 1.62 eV (2.42 eV) at the PBE (HSE06) level of theory with the valence band maximum (VBM) along the Γ to K direction and the conduction band minimum (CBM) at the Γ point. The VBM is mainly contributed by the $p_z$ orbitals of Se atoms, while the CBM by a hybridization between the $s$ orbitals of In atoms and the $p_x$, $p_y$, and $p_z$ orbitals of Se atoms (Fig. S1). For the Sb single layer, the band gap is 1.13 eV (1.66 eV) at the PBE (HSE06) level of theory with the VBM at the Γ point and the CBM along the Γ to M direction. The VBM is dominated by the $p_x$ and $p_y$ orbitals of Sb atoms, while the CBM consists of the $p_x$, $p_y$, and $p_z$ orbitals of Sb atoms (Fig. S2).

The inclusion of SOC has a significant influence on the electronic properties of these two materials. Figure 1 (b) shows the electronic band structure of monolayer InSe with SOC included. Spin splitting of the valence bands along the Γ-K direction is observed, while the bands along the Γ-M direction remain spin degenerate. Monolayer InSe has $D_{3h}$ point-group symmetry that combines $C_{3v}$ symmetry with a $\sigma_h$ reflection. Consequently, the spin polarization obeys $P_{\pm}(k_{\parallel},\theta)=[0,0,\text{sgn}(\mp\sin(3\theta))]$, where $k_{\parallel}=\sqrt{k_x^2+k_y^2}$ and $\theta=\arctan(k_y/k_x)$ is the azimuth angle of momentum $\vec{k}_{\parallel}$ with respect to the axis along the Γ-M direction[50–52]. As a result, zero spin splitting is expected for $\theta = n\pi/3$ ($n$ is an integer number), i.e. along the Γ-M direction, and full out-of-plane spin splitting for $\theta = (2n+1)\pi/6$, i.e. along the Γ-K direction, which are consistent with our calculations. The band gap of monolayer InSe decreases slightly by 0.02 eV (0.03 eV) when including relativistic effect with PBE+SOC (HSE06+SOC), as shown in Table I. The band gap of 2.39 eV predicted by HSE06+SOC is smaller than the value of about 2.9 eV measured by photoluminescence[6] but in line with other calculations using hybrid functionals[44] or the *GW* method[53]. The difference is possibly due to the fact that calculations correspond to a free-standing monolayer, while in experiment there is a different dielectric environment due to encapsulation. For monolayer Sb, the degeneracies at the top of valence band are lifted and the bands split due to the transition from orbital angular momentum states to total angular momentum states induced by the SOC in the 5*p* states of the valence bands[54], as shown in Fig. 1 (d), with an associated reduction of the band gap by 0.27 eV (0.28 eV) when using PBE+SOC (HSE06+SOC). Spin degeneracy is preserved even in the presence of SOC due to the combination of time reversal and space inversion symmetries of monolayer Sb. The band gap predicted by PBE+SOC (HSE06+SOC) is 0.86 eV (1.38 eV), in agreement with previous calculations[49,54].

Next, we consider InSe/Sb bilayer heterostructures by vertically stacking InSe and Sb single layers. Figure 2 shows the different stacking configurations AA, AB, AC, AA′, AB′, and AC′. The binding energy per unit cell is used to characterize the strength of the interface interaction, which is defined as

$$E_b = E_{InSe} + E_{Sb} - E_{InSe/Sb} \qquad (2)$$

where $E_{InSe}$, $E_{Sb}$, and $E_{InSe/Sb}$ are the total energies per unit cell of the InSe single layer, the Sb single layer, and the InSe/Sb heterostructures, respectively. The optimized lattice constant, interlayer distance, and binding energy are presented in Table II. We find that the AB and AC′ stacking configurations have similar binding energies, 0.307 eV and 0.310 eV (equivalent to 21.6 meV/Å$^2$ and 21.5 meV/Å$^2$), respectively, a value comparable to most vdW systems (around 20 meV/Å$^2$)[55]. The interlayer distance in the AC′ stacking is 2.74 Å, slightly smaller than that in the AB stacking of 2.92 Å. The binding energies for the AC and AA′ stackings are 0.299 eV and 0.291 eV (equivalent to 20.9 meV/Å$^2$ and 20.5 meV/Å$^2$), respectively. The AA and AB′ stackings are not energetically stable, also manifested in their large interface distances of about 3.82 Å. The difference between the interlayer distances of the AA- and AC′ stackings is 1.08 Å with the binding-energy difference of 0.135 eV. We note that for the antimonene bilayer the interlayer distance is 3.81 Å, which is about 30% longer than the Sb-Sb bonds[12]. A tiny charge transfer is found in the AA stacking while a large charge transfer in the AC′ stacking (Fig. S7). This coincides with their relatively large differences of interlayer distance and binding energy. The stacking configuration AC′ determined in our work is in agreement with the geometry described in a previous study of InSe/Sb heterostructures[37].

Taking the AC′-stacked InSe/Sb heterostructure as an example, we calculate its phonon dispersion using the finite displacement method and a 4×4 supercell as implemented in the Phonopy code[56] (Fig. S5). We find no imaginary frequencies in the Brillouin zone, except for a small pocket near the Γ point arising from Fourier interpolation, suggesting that the AC′-stacked InSe/Sb heterostructure is dynamically stable. Figure 3 (a) shows its electronic band structure calculated using the PBE functional without SOC. The AC′-stacked InSe/Sb heterostructure exhibits a direct band gap of 0.75 eV (1.21 eV) at the PBE (HSE06) level of theory, making it interesting for infrared optoelectronics applications. Table II shows the band gap values for other stacking configurations. We find that the direct band gap semiconducting nature of the InSe/Sb heterostructures is independent of stacking pattern. Although both InSe and Sb single layers are indirect band gap semiconductors, the vdW heterostructure formed by them is momentum-matched, thus facilitating optoelectronics device applications. The band-decomposed charge densities in Fig. 3 (c) show that the VBM is located at the Sb layer while the CBM state is distributed across the whole heterostructure, which may lead to unconventional excitonic excitations. The orbital projected band structures and density of states (Fig. S3 and Fig. S4) show that there is strong hybridization between the Se $p_z$ and Sb $p_z$ orbitals at the bottom of the conduction bands.

When SOC is turned on, strong band splitting at the bottom of the conduction bands is observed [Fig.3 (b)], which is not found in the individual layers thus implying that it is due to the interfacial interaction and symmetry breaking in the heterostructure. The band splitting resembles the Rashba spin splitting on the surface states of heavy metals[28,29]. Figure 4 shows a constant energy contour plot of the spin texture calculated in a $k_x$-$k_y$ plane centered at the Γ point and at an energy surface 0.1 eV above the crossing point of the Rashba spin split bands. The spin direction is mainly in plane and the spin texture shows outer and inner concentric circles with counterclockwise and clockwise rotating spin directions, respectively, consistent with the characteristics of the Rashba Hamiltonian[25]. The eigenvalues of the Rashba Hamiltonian in Eq. (1) are

$$E_{\pm} = \frac{\hbar^2 k_{\parallel}^2}{2m^*} \pm \alpha_R k_{\parallel} \tag{3}$$

with $\alpha_R = \frac{2E_R}{k_R}$, where $E_R$ is the Rashba energy and $k_R$ is the momentum offset. The Rashba parameter $\alpha_R$ is a measure of the strength of the Rashba effect. For the AC′ stacking, we obtain $E_R$ = 9.2 meV and $k_R$ = 0.016 Å$^{-1}$ [see Fig. 3 (d)], so $\alpha_R$ =1.15 eV Å. The values of the Rashba parameter $\alpha_R$ for the AB, AC, and AA′ stackings are 0.97 eV Å, 0.94 eV Å, and 0.93 eV Å, respectively, as shown in Table III. Additional hybrid functional calculations show that the inclusion of exact exchange through HSE06 tends to decrease the strength of the Rashba effect, which is also observed in other vdW heterostructures[24]. In particular, the Rashba parameter $\alpha_R$ decreases by 37% to 0.73 eV Å for the AC′ stacking. Selected materials and parameters characterizing the Rashba splitting are given in Table IV, and we find that the magnitude of Rashba parameter in the InSe/Sb heterostructures is comparable to that of the Au (111) and Bi (111) surfaces and one order of magnitude larger than that of the InGaAs/InAlAs interface. The value of $\alpha_R$ in the AC′-stacked InSe/Sb heterostructure is also larger than that of the SiC/MSSe (M = Mo, W) heterostructures (~0.2 eV Å)[57] and that of the GaSe/MoSe$_2$ heterostructure (0.49 eV Å)[24]. We note that the spin splitting in Au (111) surface bands reaches a maximum of 110 meV at 0.153 Å$^{-1}$, when one of the bands crosses the Fermi level. Orbital angular momentum based models have been proposed to explain the band splitting in Au (111) states, where the splitting energy is dominantly described by the atomic SOC term[35,36]. A combination of large Rashba spin splitting and a suitable band gap of about 0.9 eV at the HSE06+SOC level of theory make InSe/Sb heterostructures potentially suitable for low temperature spintronic device applications.

To elucidate the origin of the Rashba spin splitting in the InSe/Sb heterostructures, we calculate the charge density difference between the InSe/Sb heterostructure and the monolayers, which is defined as

$$\Delta\rho = \rho(InSe/Sb) - \rho(InSe) - \rho(Sb), \tag{4}$$

where $\rho(InSe/Sb)$ is the charge density of the InSe/Sb heterostructure, and $\rho(InSe)$ and $\rho(Sb)$ are the charge densities of the relaxed free-standing InSe and Sb single layers located at the positions they have in the heterostructure. Figure 5 (a) shows the charge density difference of the AC′-stacked InSe/Sb heterostructure. The interfacial interaction leads to charge redistribution with the formation of depletion and accumulation regions for electrons, which induce an out-of-plane dipole field in the heterostructure. The plane-averaged electrostatic potential (with dipole correction) in Fig. 5 (b) highlights the potential step between the respective vacuum levels of the component single layers. Kim *et al.* showed that an asymmetric charge distribution is formed due to the existence of local orbital angular momentum $\vec{L}$ in a Bloch wave function and the band splitting is dominated by the atomic SOC, $H_{SOC} = \alpha \vec{L} \cdot \vec{S}$, in the weak-SOC case such as Au(111) surface state[36]. Szary et al. found that in the Pb/MoTe$_2$ vdW heterostructure the Rashba spin splitting originates directly from the atomic SOC, with the orbital angular momentum unquenched due to broken inversion symmetry[58]. We note that there is a large spin splitting in the lowest conduction band, as shown in Fig. 3 (b). For the CBM state [Fig. 3 (c)], the charge density retains the atomic features of the Sb, Se, and In atoms. Owing to the mirror symmetry breaking in the InSe/Sb heterostructure, the $\langle L_x \rangle$ and $\langle L_y \rangle$ components can have nonzero values. It is expected that the observed spin splitting is governed by the atomic SOC[36,58,59].

It is important to study the effect of strain and external electric fields on the electronic properties of the InSe/Sb heterostructures, as these external parameters will be present and exploitable when the heterostructures are integrated in devices. We investigate strained InSe/Sb heterostructures in the AC′ stacking by applying in-plane biaxial strain ranging from -3% (compression) to 3% (elongation). The AC′ stacking is the energetically favorable configuration under strain, as shown in Fig. S8. The system retains its direct band gap semiconducting nature as the strain changes from -3% to 1 %. The band gap is 0.25 eV and 0.56 eV under -3% compressive and 1% tensile strains, respectively, calculated using PBE+SOC. However, with further increase in tensile strain, the heterostructure becomes an indirect band gap semiconductor at 2%, as shown in Fig. 6 (c). The Rashba energy $E_R$ changes monotonously from 4 meV to 13.4 meV with the strain varying from -3% to 3%, whereas the momentum offset $k_R$ stays almost constant when the strain is larger than 1%, as shown in Fig. 6 (a). The Rashba parameter $\alpha_R$ is 0.8 eV Å and 1.58 eV Å under -3% compressive and 3% tensile strains, respectively. We note that under -3% compressive and 3% tensile strains the interlayer distances are 2.96 Å and 2.53 Å, respectively, while the equilibrium geometry value is 2.74 Å. The smaller interlayer distance in the tensile strain configuration leads to stronger interfacial interaction and thus to a larger Rashba parameter. Consequently, one can effectively tune the Rashba spin splitting by varying the in-plane biaxial strain.

We investigate the role of an external electric field by applying an out-of-plane electric field ranging from -0.2 V/Å (antiparallel to the surface normal) to 0.2 V/Å (parallel to the surface normal). The Rashba energy $E_R$ and momentum offset $k_R$ change nearly linearly as a function of the electric field, as shown in Fig. 6 (b). The Rashba parameter $\alpha_R$ is 1.09 eV Å and 1.16 eV Å under -0.2 V/Å and 0.2 V/Å, respectively. Figure 6 (d) shows the electronic band structure under an electric field of -0.2 V/Å. The band gap changes little, from 0.53 eV to 0.47 eV calculated by PBE+SOC, with the electric field varying from -0.2 V/Å to 0.2 V/Å. Since biaxial strain and external electric fields are both able to tune the Rashba spin splitting, both approaches can be used to facilitate spin manipulation in spintronic applications.

## IV. Conclusion

In summary, we investigate the stacking geometry, stability, and electronic band structure of InSe/Sb heterostructures using first-principles calculations. For the lower-energy stacking configurations, the binding energy ranges from 21.6 meV/Å$^2$ to 20.5 meV/Å$^2$, comparable to most 2D van der Waals materials. We find that the InSe/Sb heterostructures have a direct band gap independent of stacking pattern. Inclusion of spin-orbit coupling yields a large Rashba spin splitting at the conduction-band edge around the Γ point, whose strength can be tuned by applying in-plane biaxial strain or an out-of-plane external electric field. Our results demonstrate InSe/Sb heterostructures are a promising platform for low temperature spintronic and optoelectronic applications.

ACKNOWLEDGMENTS

D.F. acknowledges the financial support from the National Natural Science Foundation of China (Grant No. 11604254), the Natural Science Foundation of Shaanxi Province (Grant No. 2019JQ-240), and the China Scholarship Council (No. 201906285031). S.C. acknowledges financial support from the Cambridge Trust and from the Winton Programme for the Physics of Sustainability. B.M. acknowledges support from the Gianna Angelopoulos Programme for Science, Technology, and Innovation and from the Winton Programme for the Physics of Sustainability. We also acknowledge the HPCC Platform of Xi'an Jiaotong University for providing the computing facilities.

Table I. Optimized lattice constant $a$ and indirect band gap $E_g$ (in eV) for the single layers of InSe and Sb.

| | $a$ (Å) | VBM/CBM | $E_g$-PBE/PBE-SOC | $E_g$-HSE06/HSE06-SOC | $E_g$-exp |
|---|---|---|---|---|---|
| InSe | 4.026 | Γ-M/Γ | 1.62/1.60 | 2.42/2.39 | 2.9[a] |
| Sb | 4.064 | Γ/Γ-M | 1.13/0.86 | 1.66/1.38 | |

[a]Ref.[6].

Table II. Optimized lattice constant $a$, interlayer distance $d$, binding energy $E_b$ per unit cell, and direct band gaps $E_g$ for the different stacking configurations of InSe/Sb heterostructure.

| | $a$ (Å) | $d$ (Å) | $E_b$ (eV) | $E_g$-PBE/PBE-SOC(eV) | $E_g$-HSE06/HSE06-SOC(eV) |
|---|---|---|---|---|---|
| AA | 4.039 | 3.82 | 0.175 | 0.38/0.15 | 0.92/0.65 |
| AB | 4.052 | 2.92 | 0.307 | 0.62/0.37 | 1.11/0.85 |
| AC | 4.064 | 2.90 | 0.299 | 0.67/0.42 | 1.15/0.89 |
| AA′ | 4.049 | 3.03 | 0.291 | 0.58/0.34 | 1.08/0.82 |
| AB′ | 4.039 | 3.81 | 0.177 | 0.38/0.14 | 0.92/0.65 |
| AC′ | 4.078 | 2.74 | 0.310 | 0.75/0.50 | 1.21/0.94 |

Table III. Rashba energy $E_R$, momentum offset $k_R$, and Rashba parameter $\alpha_R$ for the AB, AC, AA′, and AC′ configurations of InSe/Sb heterostructure.

|  | PBE | | | HSE06 | | |
| --- | --- | --- | --- | --- | --- | --- |
|  | $E_R$ (meV) | $k_R$ (Å$^{-1}$) | $\alpha_R$ (eV Å) | $E_R$ (meV) | $k_R$ (Å$^{-1}$) | $\alpha_R$ (eV Å) |
| AB | 6.3 | 0.013 | 0.97 | 2.7 | 0.010 | 0.54 |
| AC | 7.5 | 0.016 | 0.94 | 3.4 | 0.012 | 0.57 |
| AA′ | 5.6 | 0.012 | 0.93 | 2.5 | 0.009 | 0.56 |
| AC′ | 9.2 | 0.016 | 1.15 | 5.1 | 0.014 | 0.73 |

Table IV. Selected materials and parameters characterizing Rashba spin splitting.

|  | $E_R$ (meV) | $k_R$ (Å$^{-1}$) | $\alpha_R$ (eV Å) |
| --- | --- | --- | --- |
| **Surface** | | | |
| Au(111)[28] | 2.1 | 0.012 | 0.33 |
| Bi(111)[29] | 14 | 0.05 | 0.55 |
| InSb(110)[60] | | | 1 |
| **Interface** | | | |
| InGaAs/InAlAs[30] | <1 | 0.028 | 0.07 |
| **Bulk** | | | |
| BiTeI[31] | 100 | 0.052 | 3.8 |
| GeTe[61] | | | 5 |

**Figure captions**

Figure 1. Electronic band structures of single-layer InSe [(a) and (b)] and Sb [(c) and (d)] without and with SOC. The zero of energy is set at the valence band maximum.

Figure 2 Top view and side view of the relaxed geometries of InSe/Sb heterostructures for different stacking configurations. The primitive unit cells used in the calculations are indicated by the black solid lines.

Figure 3 Electronic band structure of the AC′-stacked InSe/Sb heterostructure: (a) without SOC and (b) with SOC. The zero of energy is set at the valence band maximum. (c) The band-decomposed charge densities for the valence band maximum (VBM) and conduction band minimum (CBM) states at the Γ point with isosurface values of 0.005 and 0.003 electron/Bohr$^3$, respectively. SOC is excluded in the charge density calculations. (d) Zoom in of the Rashba spin splitting at the bottom of conduction band.

Figure 4 Spin texture of the AC′-stacked InSe/Sb heterostructure calculated in a $k_x$-$k_y$ plane centered at the Γ point and at an energy surface 0.1 eV above the crossing point of the Rashba spin split bands. Color depicts the spin projection.

Figure 5 (a) Charge density difference due to the interfacial interaction (isosurface value: 0.0005 electron/Bohr$^3$). The cyan and yellow colors indicate the depletion and accumulation regions for electrons, respectively. (b) Plane-averaged electrostatic potential along the out-of-plane direction. The dashed lines highlight the positions of the Sb and Se atoms.

Figure 6 Left: Variations of Rashba energy and momentum offset for the AC′-stacked InSe/Sb heterostructure under (a) in-plane biaxial strains and (b) out-of-plane external

electric fields. Right: The electronic band structures under (c) 2% biaxial tensile strain and (d) -0.2 V/Å external electric field. The zero of energy is set at the valence band maximum.

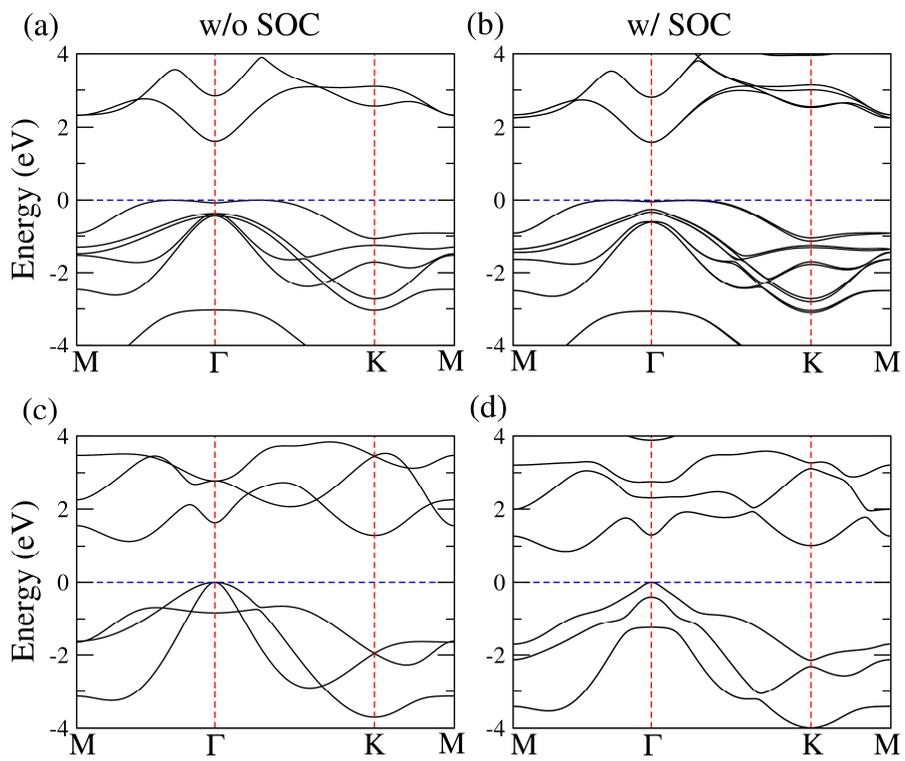

Figure 1

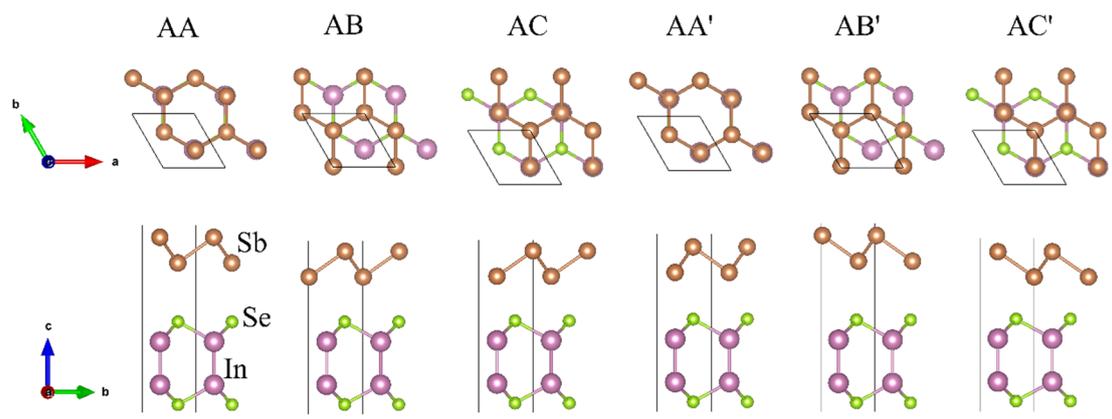

Figure 2

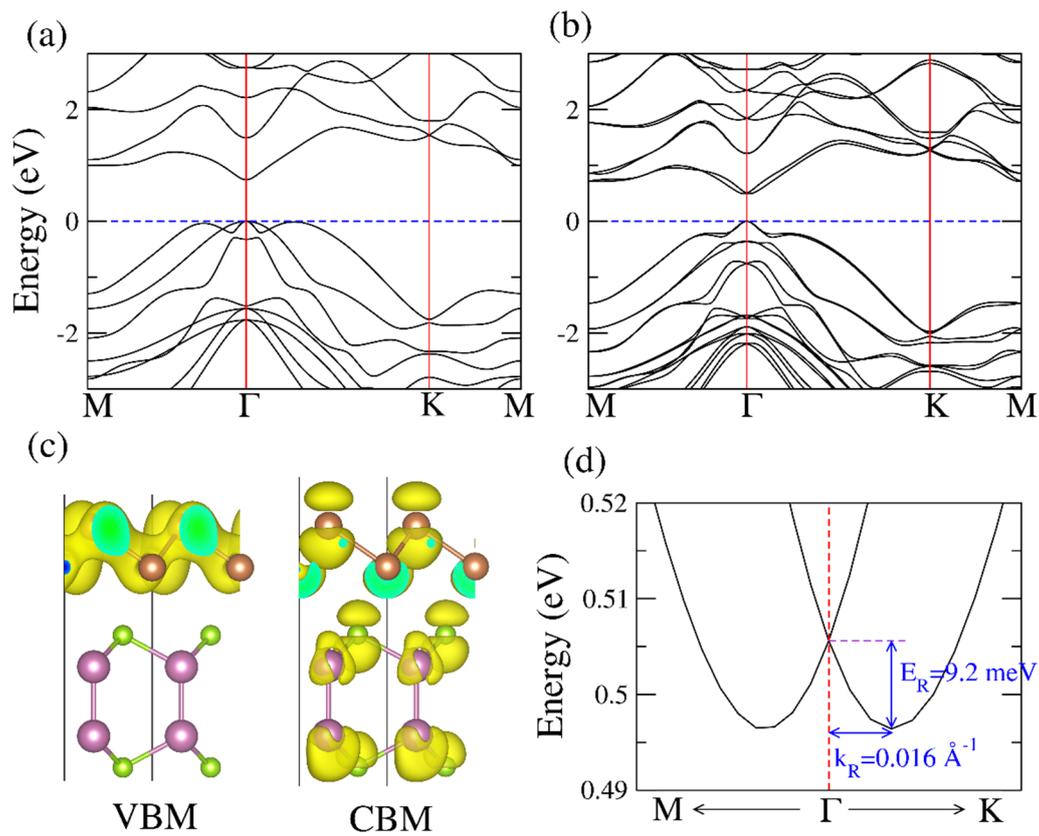

Figure 3

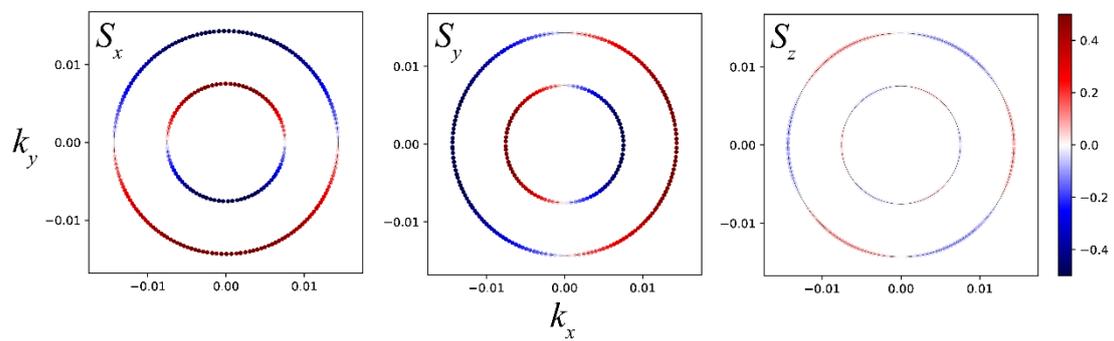

Figure 4

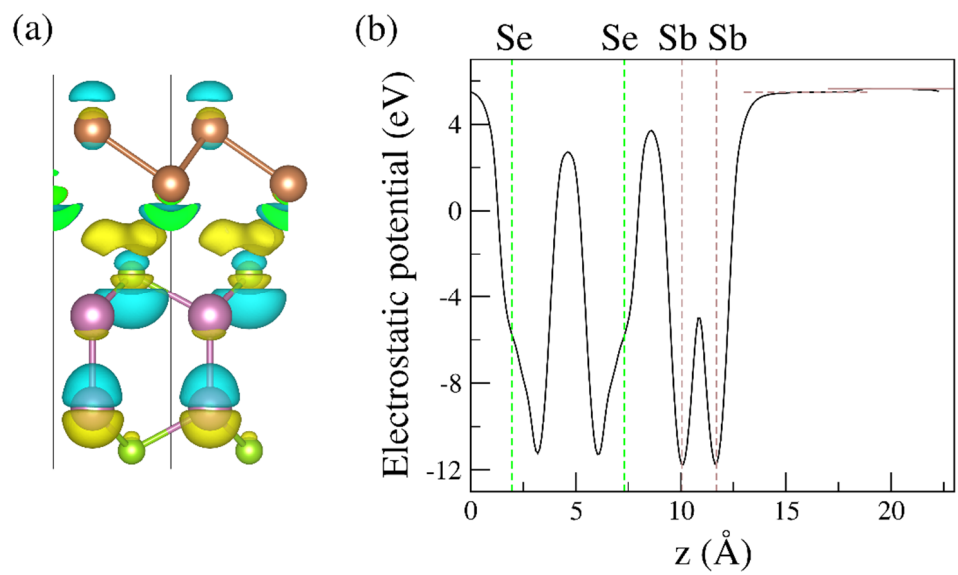

Figure 5

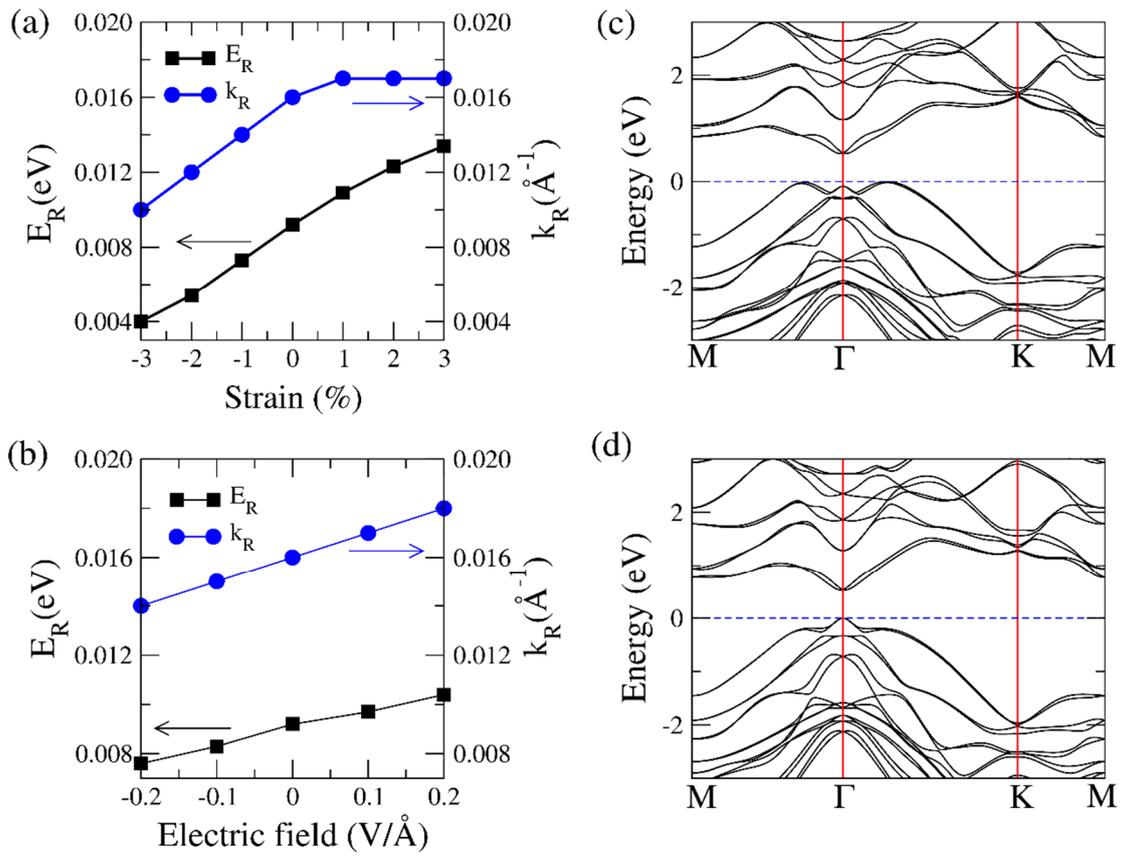

Figure 6